# Interlinking helical spin textures in nanopatterned chiral magnets

**Authors:** *Luke Alexander Turnbull,\*[1,2] Max Thomas Birch,[3] Marisel Di Pietro Martínez,[1,2] Rikako Yamamoto,[1,2] Jeffrey Neethirajan,[1] Marina Raboni Ferreira,[1,4] Elina Zhakina,[1] Hayden Jeffrey Binger,[1] Young-Gwan Choi,[1] Rachid Belkhou,[5] Simone Finizio,[6] Markus Weigand,[7] Dieter Suess,[8] Daniel Alexander Mayoh,[9] Geetha Balakrishnan,[9] Claas Abert,[7] Sebastian Wintz,[7] and Claire Donnelly\*[1,2]*

*Affiliations:*
[1] *Max Planck Institute for Chemical Physics of Solids, Nöthnitzer Str. 40, 01187 Dresden, Germany.*
[2] *International Institute for Sustainability with Knotted Chiral Meta Matter (WPI-SKCM²), Hiroshima University, 1-3-1 Kagamiyama, Higashi-Hiroshima, Hiroshima 739-8526, Japan*
[3] *RIKEN Center for Emergent Matter Science, 351-0198, Saitama, Japan*
[4] *Brazilian Synchrotron Light Laboratory, Brazilian Center for Research in Energy and Materials, 13085-970 Campinas, Brazil*
[5] *Synchrotron SOLEIL, Saint Aubin, FR-91192, France*
[6] *Swiss Light Source, Paul Scherrer Institute, 5232 Villigen PSI, Switzerland*
[7] *Institute for Nanospectroscopy, Helmholtz-Zentrum Berlin, 12489 Berlin, Germany*
[8] *Physics of Functional Materials, Faculty of Physics, University of Vienna, Kolingasse 14-16, A-1090, Vienna, Austria*
[9] *Department of Physics, University of Warwick, Coventry, CV4 7AL, UK*

\*Correspondence to: *luke.turnbull@cpfs.mpg.de, claire.donnelly@cpfs.mpg.de*

**Nanoscale topologically non-trivial magnetization configurations generate significant interest due to both the fundamental properties of their knotted structures[1], and their potential applications in ultra-efficient computing devices[2,3]. While such textures have been widely studied in two dimensions[4,5], three-dimensional (3D) systems can yield more complex configurations[6], resulting in richer topologies and dynamic behaviors[7]. However, reliably nucleating these 3D textures has proven challenging and so far, 3D configurations, such as vortex rings[8,9] and hopfions[10,11], can often only be observed forming spontaneously in relatively uncontrolled manners. Here, we demonstrate that through the 3D nanopatterning of chiral single crystal helimagnets into nano-tori, the controlled formation of a magnetic double helix can be achieved. This surface localized topological state is stabilized by the interplay of intrinsic exchange interactions of the single crystal with the extrinsic emergent effects of the patterned geometry. These double helices host magnetic defects akin to supercoiling in circular DNA[12] and climbing vines[13]. We expect this study to serve as a foundation for future research combining single crystal systems with 3D nanopatterning, offering a new degree of control over emergent phenomena in nanoscale magnets and wider quantum material systems.**

Broken symmetries drive complex magnetic phenomena, from non-reciprocal dynamics[14] to the formation of topologically non-trivial textures[15]. In chiral helimagnets, broken inversion symmetry in the unit cell induces an antisymmetric exchange interaction, commonly known as the Dzyaloshinskii-Moriya interaction (DMI)[16,17]. In competition with the direct exchange interaction, the DMI promotes the formation of a wide range of topological solitons, such as skyrmion tubes[18] and hopfions[19]. However, the localized and controlled nucleation of such 3D textures remains a significant challenge. An alternative route to break the symmetry of a system involves the patterning of nanoscale 3D geometries, where the introduction of geometric confinement, curvature, and torsion have been exploited to manipulate topological defects[20], as well as induce chirality in otherwise achiral magnetic materials[15,21]. Until now, 3D geometries have mainly been combined with deposited ferromagnetic systems that typically host relatively uniform magnetic states. In these systems, the geometric symmetry breaking has led to phenomena that would more commonly be associated with more complex materials[22]. Until now the combination of intrinsic and extrinsic symmetry breaking has not been explored.

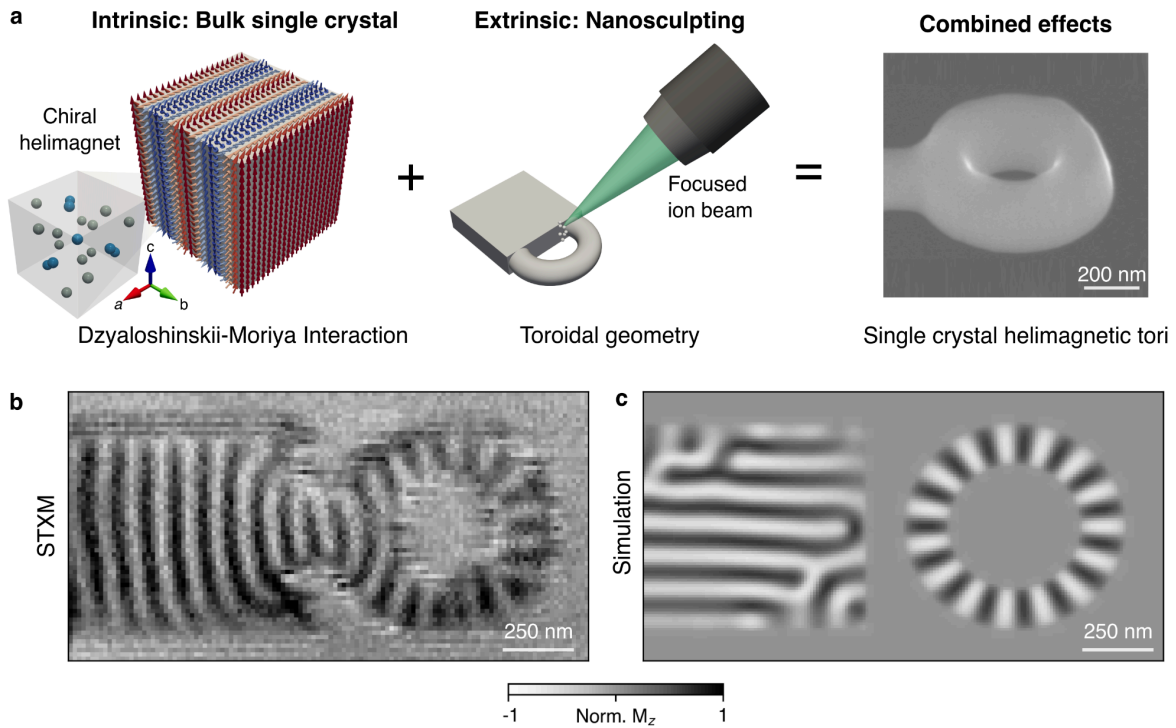

**Figure 1: Combining single crystal helimagnets with 3D nanosculpting. a)** The combination of intrinsic magnetic interactions arising from the symmetries of a single crystal, with extrinsic effects arising from a 3D nanosculpted geometry is a potential route for tailoring the magnetic energy landscape of a system. The scanning electron micrograph reveals a nanoscale torus formed by milling a single crystal helimagnet, $Co_8Zn_9Mn_3$. **b)** The magnetic state of a slab-like lamella and torus is probed using STXM and XMCD, providing a projection of the magnetization out of the plane. The image reveals the magnetic ordering in the torus is distinct from the straighter helical stripes in the lamella, due to the radial ordering of the magnetic stripes. **c)** Finite element micromagnetic simulations of an equivalent geometry reveals similar projections of the respective states. Notably there is a wedging in the torus magnetization stripes.

Here, we achieve a 3D topologically non-trivial magnetic texture by combining intrinsic and geometric symmetry breaking in a helimagnetic nano-torus (Fig. 1a). The nanoscale geometry introduces confinement, curvature, and non-trivial topology to the system. In this way, we combine two forms of symmetry breaking: global, through the broken inversion symmetry of the chiral unit cell in the helimagnet, and local, through the patterning of a curved geometry, to realize a new configuration – a magnetic double helix state - analogous to a spring-like configuration. The texture can deform within the sample, and exhibits defects similar to those in non-magnetic helical systems, such as negative supercoiling defects in DNA. The reproducible stabilization of this chiral 3D topological state demonstrates the potential for realizing otherwise inaccessible phenomena in nanopatterned single crystal systems.

To nanosculpt bulk single crystals of the room-temperature helimagnet $Co_8Zn_9Mn_3$, we harness the micropatterning capabilities of focused ion beam milling[23,24]. The curved geometry is created by exploiting selective milling patterns, and the Gaussian profile of the ion beam[25]. An example of a fabricated nano-torus is shown in Fig. 1a. Chiral helimagnets display an archetypal helical rotation of the magnetization vector, due to the competition of the direct exchange interaction, and DMI. In $Co_8Zn_9Mn_3$, this rotation has a characteristic pitch $\lambda = 120$ nm[26], set by the relative strength of these two exchange interactions. Therefore, we aim to pattern the geometry to length scales approaching this characteristic length of the system[27].

We determine the effect of the 3D nanopatterning on the resulting magnetization configuration of the helimagnet by performing high spatial resolution X-ray microscopy (see Methods). The main effects are illustrated in a sample with two connected regions: a lamellar region and a toroidal region that combines a confined circular cross section, with a closed loop curved geometry. The out-of-plane magnetization projection (Figure 1b) shows a striking difference in the two magnetic configurations. In the lamella, straight stripes appear with a uniform periodicity matching the characteristic helical period of $\lambda = 120$ nm in $Co_8Zn_9Mn_3$[26]. However, in the toroidal section of the same sample, the stripes propagate azimuthally around the torus, indicating a geometrical modification to the magnetic ordering of the helimagnet. In confined 2D systems it is common to observe magnetic helices propagating tangentially to the sample boundary[28–30]; this behavior is replicated here, with the curvature of the torus ring also causing a constant rotation of the helical wavevector, and an associated wedging of the radial stripes.

One notable property of emergent curvilinear effects in magnetism is that they can arise from exchange interactions, in contrast to traditional shape anisotropy effects, which are magnetostatic in origin and form the basis of almost all magnetic geometric effects[27,31]. We determine the origin of our geometric effect by performing finite element micromagnetic simulations of a torus of similar dimensions (Fig. 1c). Remarkably, we are able to reproduce the state even in the absence of magnetostatics, considering only the direct exchange interaction and DMI (an analysis of the role of magnetostatics is provided in the Supplementary information). This indicates the interplay of exchange and DM interactions within the nano-torus geometry is sufficient to recover the observed behavior of the system.

To determine the influence of the geometry of the nano-torus, we next consider structures of varying dimensions, both experimentally and numerically (Fig. 2). Each nano-torus can be characterized by two radii: the torus radius, $R_T$, and the nanowire cross-sectional radius, $r_{NW}$. By performing X-ray microscopy measurements on a range of different toroidal geometries, we identify two ordering regimes: an radial ordering of helical stripes around the torus, and a bulk-like regime where the stripes remain straight, and are able to form multiple windings across the cross-sectional diameter. Figure 2(a,b) shows representative images of these two regimes, and projections of the corresponding micromagnetic simulations are displayed below in Figure 2(c,d) (images of the full set of tori with different geometric parameters are available in the Supplementary information). We plot a geometric phase diagram of the helical state in Figure 2e where one can observe good agreement between our measurements and simulations. From this phase diagram it is apparent that the radial ordering state is entirely dependent on confinement due to the nanowire radius, $r_{NW}$, with the state emerging when $r_{NW} \lesssim 1.2 \lambda$; above this critical radius the bulk-like state is stabilized. The limiting regime of $r_{NW} \lesssim \lambda/2$ was not realized experimentally and is addressed in the Supplementary information.

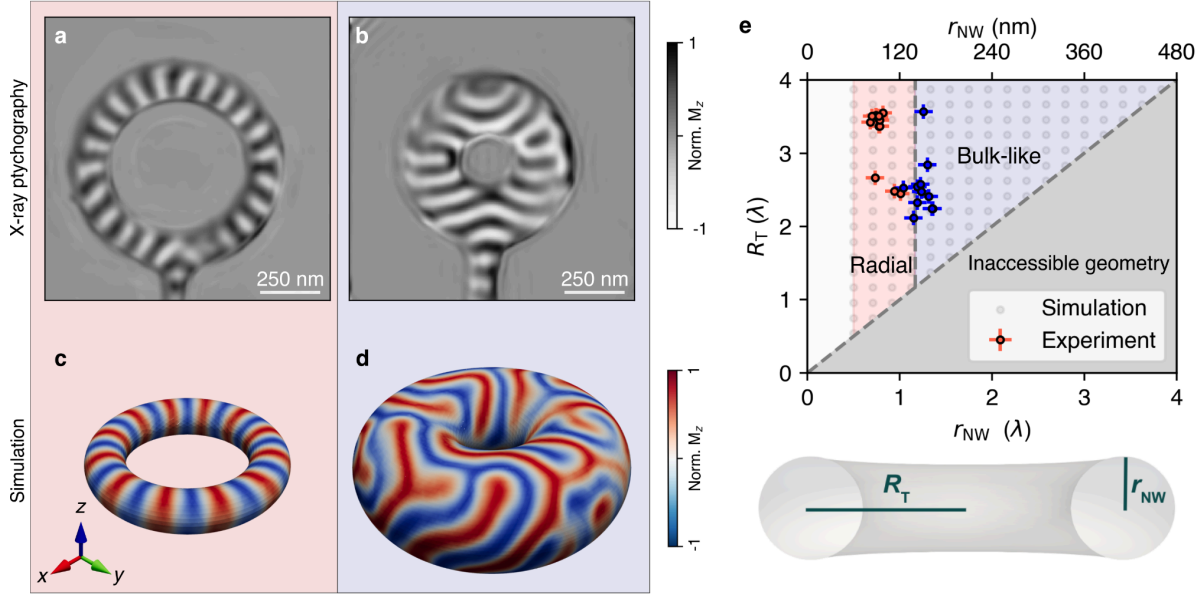

**Figure 2: Geometry dependence of the magnetic state within toroidal helimangets. a)** XMCD image of a torus ($R_T$ = 500 nm, $r_{NW}$ = 120 nm) with the radial ordering of the magnetic stripes. **b)** XMCD image of a torus ($R_T$ = 380 nm, $r_{NW}$ = 260 nm), with a more disordered bulk-like ordering. **c,d)** Finite element micromagnetic simulations of tori with similar minor radii ($R_T$ = 380 nm, $r_{NW}$ = 90 nm; $R_T$ = 400 nm, $r_{NW}$ = 260 nm). **e)** Experimental and simulated geometric phase diagram for the helimagnetic tori. States identified experimentally with XMCD imaging are indicated by '+'s, while the simulated states are indicated by the circles and color scale. There are two distinct ordering regimes established, dominated by the minor radius of the torus. The red and blue regions mark the radial ordering, and bulk-like ordering respectively.

We delve further into the role of confinement by considering the full 3D magnetic configuration of the simulated nano-tori, in particular the ordering at the surface (Fig. 3a). When examining a series of slices spanning a single helical period around the circumference of the torus (Fig. 3b), we observe that within the center of each cross-section, the magnetization is uniform, exhibiting the characteristic helical rotation between each respective slice. This behavior is driven by the intrinsic DMI of the helimagnet system, as expected within the bulk. However, when plotting the magnetization out of the plane of each slice, we observe an anisotropic canting of the magnetization in the vicinity of the surface which is maximal when the magnetization is locally orthogonal to the surface normal. The canting profile can be seen when a line profile of the magnetization across the slice is plotted (Fig. 3c), where it is apparent there is a canting decay length of ~ 25 nm. Displayed as a function of $r_{NW}$, we see that the canting consistently relaxes over a length of ~ 25 nm into the bulk, indicating that it is indeed a localized surface effect, and not due to bulk helical winding (Fig. 3c). This behavior is a manifestation of the previously observed edge twists in bulk and 2D helimagnetic systems[32,33]. However, here with a 3D curved surface, we observe this edge twisting leads to a complex surface-induced texture.

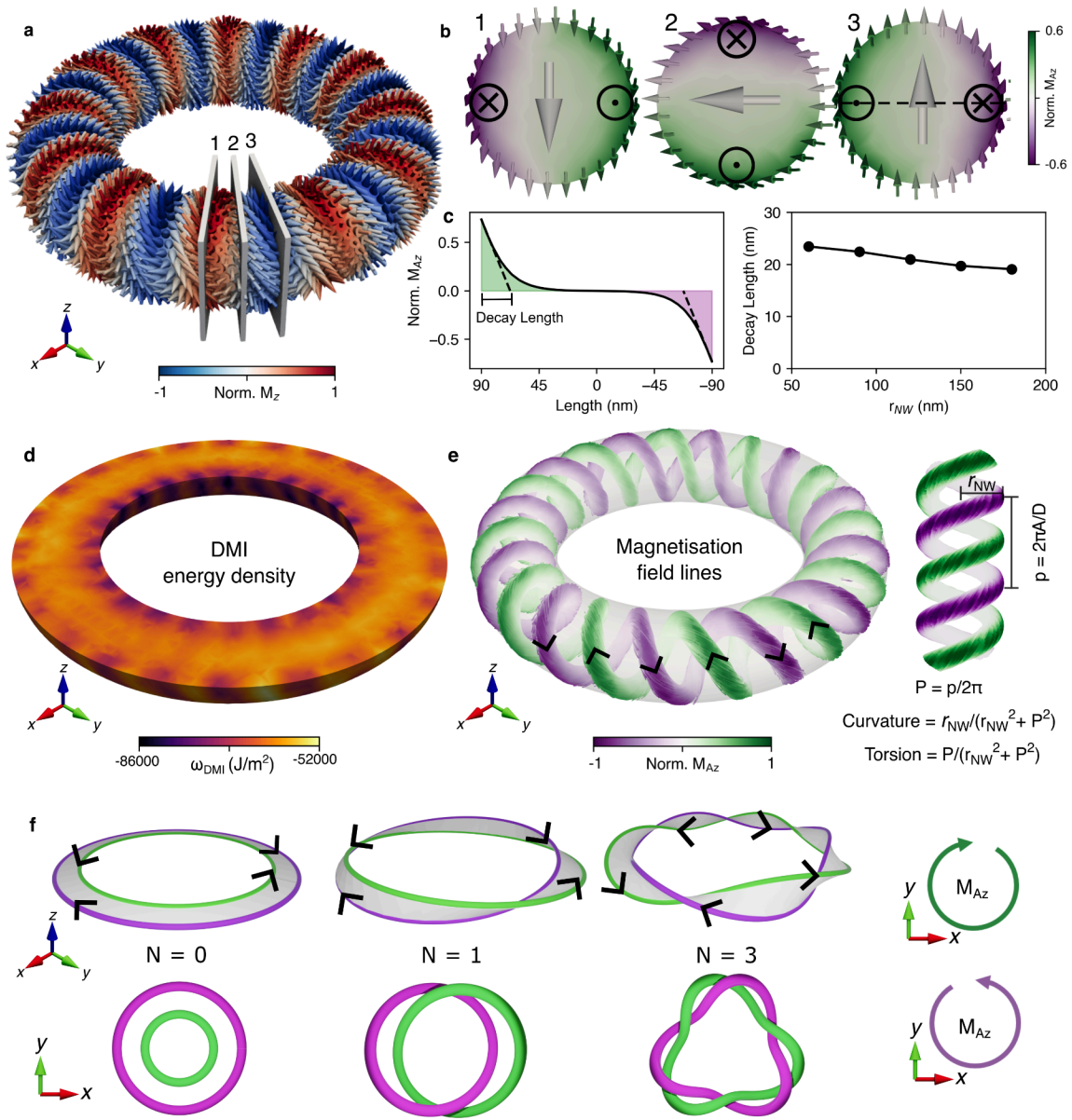

**Figure 3.** Formation of magnetic double helix texture due to curved toroidal surface with micromagnetic simulations. a) Rendering of the surface magnetization reveals the azimuthal canting of the magnetization. b) Magnetization in slices 1-3 over three quarters of a single helical period (defined in a). In each slice, the volume of the torus points in a single direction, while there is an anisotropic canting of the magnetization at the surface, which depends on the local orientation of the surface normal and magnetization. c) The magnitude of the canting can be seen when a line profile of the magnetization across slice 3 is plotted, giving a canting decay length of 25 nm. Here, $r_{NW}$ is 90 nm, and $R_T$ is 360 nm. The canting is maximal when the magnetization and surface normal are perpendicular. The canting decay length as a function of nanowire radius for different tori, is relatively constant and relaxes to a value of 25 nm approaching the bulk-like regime. d) Rendering of the DMI energy density for a cut through the plane of the torus. There is a periodic reduction of the DMI energy density at the surface of the torus, corresponding to the regions of maximal canting. e) Field lines of the canted surface magnetization that indicate the local tangent, or "flow" of the magnetization vector field reveal

two geometric helices with locally opposite orientation of the magnetization, i.e. an antiparallel double helix texture. The pitch, *p*, of this helix is determined by the intrinsic interactions, while the curvature and torsion are determined by a combination of the intrinsic interactions and patterned geometry. *A* and *D* are the strengths of the direct exchange and DM interactions respectively. f) We consider the topology of the double helix texture by considering the evolution of the orientation of the magnetization as we traverse around the torus. We plot a surface to which the magnetization is normal, which forms a closed ribbon with a number of twists corresponding to the number of periods of the volume helix. The edges of this ribbon correspond to the regions of maximal canting, which twist and therefore form loops that link one another. This twisting or linking number is an integer and quantifies the topology of the magnetic configuration.

To understand the surface-induced canting, we consider the magnetization within a single slice of the torus. In the bulk, DMI induces the characteristic rotation between adjacent slices, manifesting as a one-dimensional modulation, without any canting in the planes normal to the modulation direction. The one dimensional nature of this helical winding occurs as introducing additional canting within these planes locally increases the exchange energy density. At the same time, the symmetry of the DMI requires that any winding embedded within a uniform background must complete a full rotation of the spins, otherwise there is no net gain in DMI energy. As a result, within the bulk, uniform ordering within planes perpendicular to the helices is promoted. However, at the sample surface, where the magnetization is no longer embedded within a uniform field, this restriction is lifted and the surface magnetization is not only free to cant, but the DMI enforces a canting. Indeed, one can see this reduction in the DMI energy density in the vicinity of the canting in Figure 3(d), where there is a periodic lowering of the energy density, as compared to within the volume of the nanowire.

The nature and extent of this canting can be formulated by considering the impact of the surface on the local symmetries of the system. While the broken inversion symmetry of the unit cell gives rise to bulk DMI, the surface introduces additional symmetry breaking that modifies the behavior of both the exchange and DM interactions. Magnetostatic effects are neglected here, as justified by our prior simulations, though notably their influence at a sample boundary is not explicitly modified by the additional symmetry breaking. By treating the magnetization, ***m***, as a continuous vector field, as in micromagnetics, we can represent the surface behavior of a helimagnet as[34–36]:

$$D(\bm{n} \times \bm{m}) = A(\nabla \bm{m} \cdot \bm{n}), \quad (1)$$

where ***n*** is the vector normal to the sample surface, and *D* and *A* represent the strengths of the DM and exchange interactions, respectively. The DMI term at the surface dictates a canting of the magnetization about the surface normal, while the exchange term scales this effect based on the orientation of the magnetization relative to the sample surface[32]. Together, they describe a canting in the magnetization about ***n***, which vanishes for magnetization parallel to the surface normal and maximal for magnetization orthogonal to the surface; this is precisely the anisotropic canting we observe in our simulations.

Having determined the mechanism of the surface induced canting, we next explore the influence of the 3D curved surface. Indeed, the combination of the rotating volume magnetization and the smooth curved surface leads to a continuously canted winding at the surface of the torus. This continuous canting leads to a 3D surface state, revealed by plotting the field lines of the magnetization, shown in Figure 3e. These field lines represent continuous flow of the magnetization, analogous to the trajectory of particles in a velocity field. When we consider the field lines, remarkably, we observe the presence of two distinct strands of continuous magnetization that wrap around the surface, forming a surface-localised magnetic double helix. Each strand propagates azimuthally around the torus, forming closed loops that circulate antiparallel to the other, reminiscent of the ground state antiparallel configuration of patterned double helix nanostructures with geometric chirality[15,21]. However, here, the geometric topology of the torus alters the connectedness of the system, distinct from what we would observe in an open-ended nanowire, or bulk sample. The parameters of our magnetic double helix are determined by the combination of both the intrinsic interactions and the geometry. The winding length determines the pitch of the double helix, while its radius is determined by the geometric radius of the nanowire, meaning that the curvature and torsion of this configuration can be tuned[37,38]. Since this configuration arises purely from the competition between helical winding and the curved surface, this 3D chiral structure can be realized in any helimagnet that is patterned on relevant length scales, defined by the bulk helical winding length.

The role of topology becomes particularly clear when we consider the number of helical windings within the texture. In a straight nanowire the magnetic orientation at each end is not fixed, and is therefore free to rotate, allowing for a smooth change in the number of helical winds. However, in the closed loop of the torus, an integer number of winds is required to retain a smoothly varying magnetization, meaning that, in the absence of defects, the helical winding is quantized. We can therefore define a linking number, $N$, for the surface helices, which serves as the topological index of the magnetic configuration and corresponds to the number of full helical windings within the torus (Fig. 3f). States with different topological indices cannot be smoothly deformed into each other; their transformation requires the mediation of a topological defect or discontinuity. Such closed-loop linking is a topic of interest in knot theory, and is relevant to circular DNA and RNA, as well as some macromolecular catenanes[39] and entangled polymers[40].

Although there is a well defined optimal linking number for tori of a certain radius, in our experiments we frequently observed a number of cases of lower linking numbers, accompanied by apparent defects in the helical ordering. Indeed, when we experimentally examine the nano-tori, we observe the formation of states containing 'U' shaped defects in the X-ray microscopy projections, outlined by squares in Fig. 4 (a-b). These defects disrupt the radial stripes, and correspond to a reduced number of stripes in the system. We are able to reproduce these defects in our micromagnetic simulations by relaxing from randomized initial states, indicating that these are metastable states of the torus (Figure 4c). Quantitatively, our simulations show that introducing a single defect increases the global energy by only ~ 0.3%, confirming that these states are near degenerate to the defect-free double helix and therefore readily stabilized. We can understand the nature of these defects from the point of view of the double helix surface state by isolating the field lines localized at these defects (Fig. 4d), where we now observe a deformation of the helical strands: an elongated inner strand and a tightly coiled outer. Significantly, there is no linking occurring in the vicinity of the defect, meaning that the presence of the defect reduces the total linking number of the system.

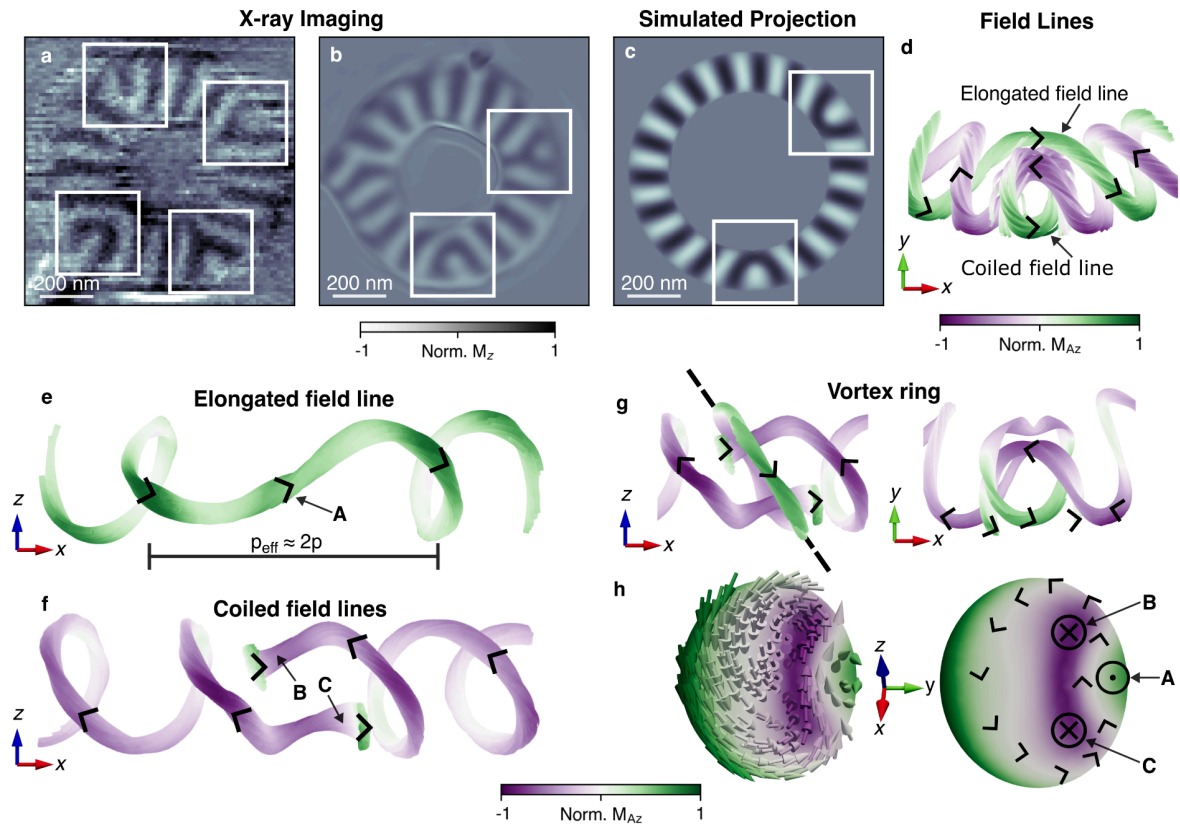

**Figure 4. Supercoiling defects in the magnetization.** a-b) X-ray imaging data of two tori, initialized by saturating the samples applying a 200 mT out of the torus plane before removing the field at room temperature, reveals the formation of two defects in the double helix state, namely "U" shaped disruptions to the radial stripes. c-d) Micromagnetic simulation relaxed from randomized initial states, also readily to the formation of the same shape defects. The associated field lines reveal an elongated field line, and a coiled field line that do not link each other. e-f) Isolated renderings of the elongated and coiled field lines reveal their respective 3D structures. The elongated field line exhibits an approximate doubling of its effective pitch. The coiled field lines are split and in the vicinity of the defect begin to turn against their natural circulation and are ejected from the sample surface. g) Renderings from two perspectives of the coiled field lines, including the vortex ring that circulates around the split strands. h) A slice of the magnetization taken from the plane indicated in panel g, through which the vortex ring circulates. In the schematic view, on the right, the circulation of the vortex ring is indicated, as well as the intersection of the extended field lines with this plane. Points A, B, C correspond to the intersection of the elongated and coiled field lines with the displayed plane respectively. These locations are marked at the equivalent positions in panels e-f.

The structure of these defects becomes clearer when examining the field lines individually. We first consider the elongated field line in Figure 4e. In the vicinity of the defect, the helical pitch is increased, with an associated decrease in the curvature and torsion. Notably, the pitch is approximately doubled, indicating that the defect replaces a single link, reducing the linking number by one. In contrast to the elongated strand, which remains continuous, we find the tightly coiled strand is split. Close to the surface, the two coils straighten and turn against their natural chirality, ultimately terminating at the sample surface, at the green ends seen in Figure 4f. This discontinuity of the field lines breaks the non-trivial topology of the linked state by ejecting the field lines from the sample surface. Despite the metastable nature of this state, we do not observe relaxation into the closed double helix ground state on experimentally accessible timescales. This behavior suggests the existence of a significant energy barrier separating the defect-containing and defect-free configurations, as expected for topologically distinct states.

While the surface canting associated with the missing wind of the double helix is lost, an alternative surface canting in the form of a vortex can be identified by a closed loop of the field lines in the vicinity of the defect (Fig. 4g). Indeed, when we plot the magnetization in the plane of this closed loop, we observe a well defined circulating magnetization with a central core that points out of the plane (Fig. 4h). As opposed to the surface localized double helix state, this vortex core propagates through the bulk of the torus, disrupting the central one dimensional helical winding. Analogous defects and coiling occur in, for example, climbing plants and circular DNA, where so-called supercoiling forms to reduce the stress associated with underwinding[13,41,42]. However, in such systems, deformations are limited to geometric deformations of the strands themselves, which form the entire structure. Here, as our double helix is embedded in a continuous medium, the underwinding forces the formation of additional topological defects, akin to turbulent vortices in velocity fields.

In conclusion, we have demonstrated that the 3D nanopatterning of single crystal helimagnets provides a pathway to define topologically non-trivial 3D magnetization configurations. By patterning nano-tori, we have combined confinement, curvature, and geometric topology on the nanoscale, reshaping the magnetic configuration and the magnetic energy landscape. The competition between intrinsic exchange and DM interactions, and the geometry-induced effects give rise to a surface-localized double helix. This three dimensional magnetic configuration is topologically non-trivial, and hosts defects akin to supercoiling in plants and DNA, which nucleate additional localized topological textures, establishing a new solid-state platform to explore the physics of interlinked configurations[43].

The double helix texture presents an opportunity for exploring knotted and interlinked configurations on the nanoscale, with implications for both device applications and the study of fundamental magnetism. Although here we consider a single torus, the extension to higher genus structures offers a route to stabilizing more complex knotted textures[44]. Beyond the zero field state, we expect the competition of intrinsic and extrinsic effects to strongly influence the statics and dynamics of field-induced textures such as skyrmions[45,46] and hopfions[19], offering a route to their controlled formation and manipulation.

Going forward, the ability to sculpt arbitrary 3D geometries from single crystals established a general framework for tailoring the interplay of intrinsic interactions with extrinsic curvature. In magnetism, this approach may provide new routes to stabilize and manipulate skymions, hopfions, and related textures. Beyond magnetism, the method can be applied to other ferroic and correlated systems, where symmetry breaking and curvature influence collective states. For example, introducing symmetry breaking into ferroelectrics, where surface effects and strain gradients dominate[47], offers new routes to tailor both configurations and manipulation of states, while the 3D nanopatterning of superconductors opens the possibility for local control of vortex states[48,49]. Finally, the use of helium ion FIB offers the possibility to increase the spatial resolution, and obtain even higher degrees of confinement that are needed to explore quantum phenomena[50].

**Methods:**

*Sample synthesis.*

Crystals of $Co_8Zn_9Mn_3$ were grown via the Bridgman method. Stoichiometric amounts of Co powder (Alfa-Aesar, 99.99%), Zn powder (SigmaAldrich, 99.995%) and Mn pieces (Alfa-Aesar, 99.99%) were ground together and transferred to an alumina crucible with a pointed end, and sealed inside an evacuated quartz tube. The tube was then heated to 1060 ℃ and allowed to homogenize for 12 hours. It was then slowly cooled at a rate of 1 ℃/hour to 700 ℃ and left to anneal for several days at this temperature before being water quenched. Single crystals of $Co_8Zn_9Mn_3$ were isolated from the as-grown boule and oriented using X-ray Laue back-reflection.

*Focused ion beam fabrication.*

Using a gallium focused ion beam, thin lamellae of $Co_8Zn_9Mn_3$ were extracted from the single crystal using a typical lift-out method. These lamellae were attached to Cu transmission electron microscopy grids, with a local Pt deposition. The toroidal samples were then shaped with targeted milling patterns, with a potential of 16 kV, before being transferred onto $Si_3N_4$ substrates with a micromanipulator, and secured in place with a local Pt deposition.

*X-ray microscopy.*

*Scanning Transmission X-ray Microscopy (STXM)*

Scanning transmission X-ray microscopy measurements were carried out at the MAXYMUS endstation at the BESSY-II electron storage ring operated by the Helmholtz-Zentrum Berlin für Materialien und Energie. Using a Fresnel zone plate and order selecting aperture, the X-ray beam was focused to a spot size of ~ 20 nm. To acquire an image, the sample was scanned through the X-ray beam using a piezoelectric motor stage, with the transmission measured pixel by pixel using an avalanche photodiode. Magnetic contrast parallel with the beam was achieved by tuning the X-ray energy to the Co $L_3$ edge (779 eV), exploiting the resonant absorption effects of X-ray magnetic circular dichroism (XMCD). All presented images are the result of subtracting two images acquired with positive and negative X-ray circular polarization, eliminating all topographic contrast and leaving only the XMCD magnetic contrast. The pixel size of the STXM images is 20 nm.

*X-ray Ptychography*

X-ray ptychography measurements were carried out at the HERMES endstation at the SOLEIL synchrotron. Using a Fresnel zone plate a coherent probe beam with a spot size of ~50 nm was produced. To acquire an image, the sample was scanned through the X-ray focal point using a piezoelectric motor stage, with the coherent scattering measured on an 2048 x 2048 pixel$^2$ area detector. Magnetic images containing phase and amplitude information were reconstructed using the PyNX platform. Magnetic contrast parallel with the beam was achieved by tuning the X-ray energy to the Co $L_3$ edge (779 eV), exploiting the resonant absorption effects of X-ray magnetic circular dichroism (XMCD). All presented images are the result of subtracting two phase images acquired with positive and negative X-ray circular polarization, eliminating all topographic contrast and leaving only the XMCD magnetic contrast. The resolution of ptychography images is set by the scattering collected on the detector. In the images presented in this manuscript the reconstructed pixel size is 8 nm.

*Finite element micromagnetic simulations.*

Micromagnetic simulations were performed with the finite-element method using the magnum.pi software. Simulations of the nano-tori were performed considering exchange and bulk DMI interactions. The exchange and DMI constants used were $A = 5.7$ pJ/m and $D = 0.6$ mJ/m$^2$, respectively[26]. The saturation magnetization was $M_s$ = 230 kA/m. Here the bulk DMI energy density takes the form: $\omega = Dm \cdot (\nabla \times m)$. The nanotori were initialized with random starting configurations, and the energy of the system was minimized by allowing the system to relax following the Landau-Lifshitz-Gilbert equation over 20 ns. The Gilbert damping parameter was set to α = 1. Simulations considering the demagnetising field are discussed in the supplementary information.

**Acknowledgements**
We acknowledge SOLEIL for provision of synchrotron radiation facilities. We thank the Helmholtz-Zentrum Berlin für Materialien und Energie for the allocation of synchrotron radiation beamtime. We acknowledge the Paul Scherrer Institut, Villigen, Switzerland for provision of synchrotron radiation beamtime at beamline Pollux of the SLS. L.A.T. and M.D.P.M. acknowledge the support of the Alexander von Humboldt Foundation. L.A.T., M.D.P.M., J.N., R.Y., E.Z. and C.D. acknowledge funding from the Max Planck Society Lise Meitner Excellence Program and funding from the European Research Council (ERC) under the ERC Starting Grant No. 3DNANOQUANT 101116043. J.N. acknowledges support from the International Max Planck Research School for Chemistry and Physics of Quantum Materials. M.R.F acknowledges support from the Max Planck Partner Group R. D. dos Reis of the MPI for Chemical Physics of Solids. The work at the University of Warwick was supported by EPSRC, UK through Grants EP/T005963/1 and EP/N032128/1.

**Author Contributions**
L.A.T., M.T.B., and C.D. conceived the project. D.A.M. and G.B. fabricated the bulk single crystal sample. L.A.T. and M.T.B. fabricated the toroidal nanostructures with a focused ion beam. L.A.T., M.T.B., M.D.P.M., R.Y., J.N., M.R.F., E.Z., H.J.B, Y.-G.C., M.W., R.B., M.W., S.F., S.W., and C.D. performed the X-ray imaging measurements. L.A.T. carried out the micromagnetic simulations, in collaboration with C.A., D.S. and C.D. The manuscript was written by L.A.T. and C.D., with input from all authors. All authors discussed the results and gave feedback on the manuscript.

**Competing Interests**
The authors declare no competing interests.

**Additional Information**
Supplementary Information is available for this paper.
Correspondence and requests for materials should be addressed to: luke.turnbull@cpfs.mpg.de.

**Data/Code Availability**
All data and code associated with this manuscript is available on request and will be made available on a repository.

**Supplementary Information: Interlinking helical spin textures in nanopatterned chiral magnets**

**Supplementary Note 1: Further experimental images of tori**

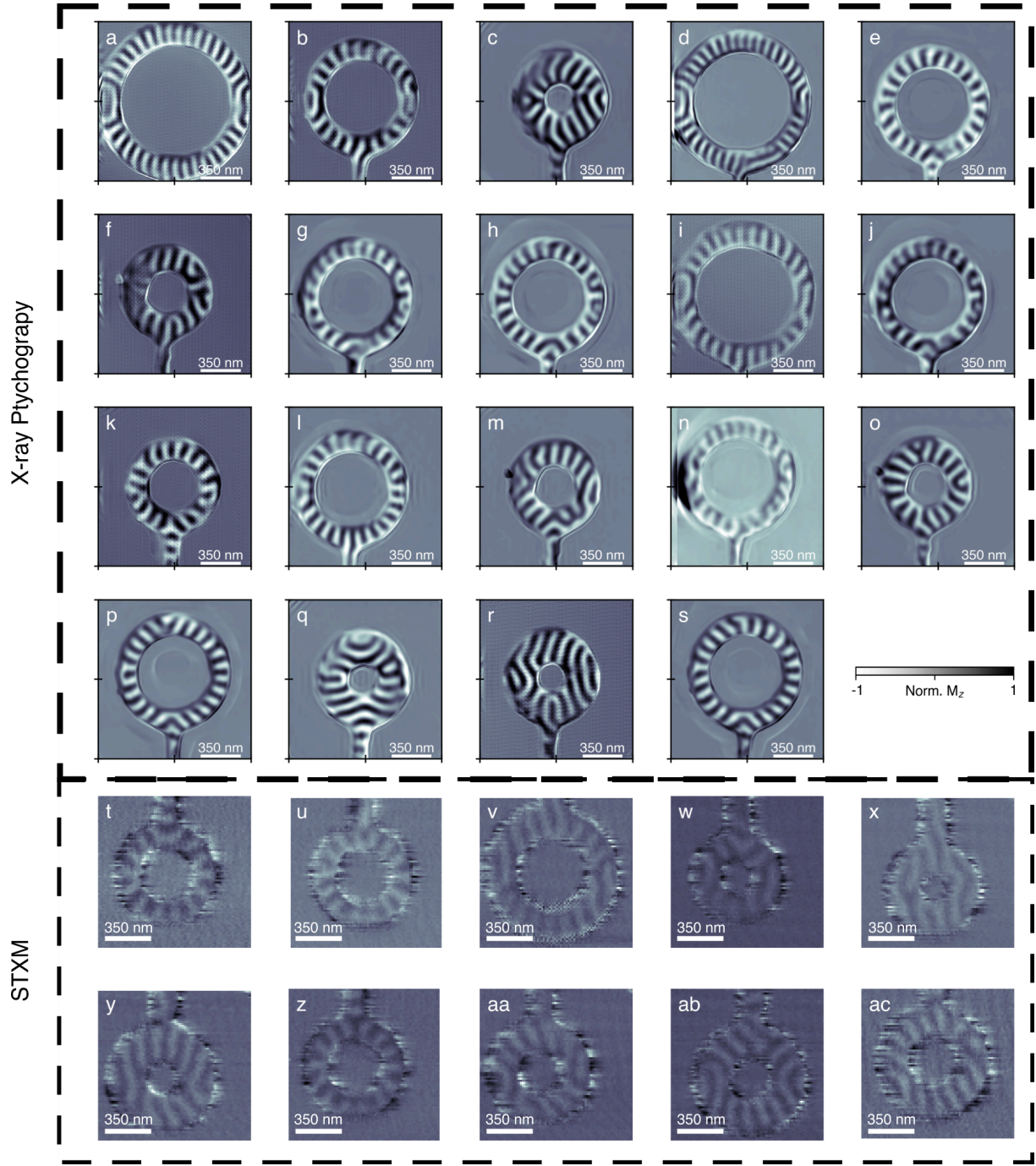

Figure S1: XMCD projections of toroidal samples. (a–s) X-ray ptychography phase images of X-ray magnetic circular dichroism (XMCD) contrast, acquired at the Co $L_3$ edge of the magnetization configuration within tori samples in a zero-field state. (t-ac) Scanning transmission x-ray microscopy images of X-ray magnetic circular dichroism (XMCD) contrast, acquired at the Co $L_3$ edge of the magnetization configuration within tori samples in a zero-field state.

| Sample Frame | $R_T$ (nm) | $r_{NW}$ (nm) | Sample Frame | $R_T$ (nm) | $r_{NW}$ (nm) |
|---|---|---|---|---|---|
| **a** | 600 | 100 | **k** | 320 | 90 |
| **b** | 410 | 80 | **l** | 410 | 90 |
| **c** | 270 | 160 | **m** | 300 | 120 |
| **d** | 530 | 90 | **n** | 430 | 100 |
| **e** | 420 | 90 | **o** | 300 | 110 |
| **f** | 290 | 120 | **p** | 420 | 80 |
| **g** | 420 | 90 | **q** | 280 | 140 |
| **h** | 400 | 90 | **r** | 290 | 160 |
| **i** | 540 | 100 | **s** | 420 | 90 |
| **j** | 400 | 90 | - | - | - |

**Table 1: Measured torus sizes from the X-ray ptychography XMCD projections in Figure S1.**

| Sample Frame | $R_T$ (nm) | $r_{NW}$ (nm) | Sample Frame | $R_T$ (nm) | $r_{NW}$ (nm) |
|---|---|---|---|---|---|
| **t** | 250 | 90 | **y** | 240 | 140 |
| **u** | 260 | 90 | **z** | 260 | 80 |
| **v** | 350 | 120 | **aa** | 240 | 110 |
| **w** | 230 | 120 | **ab** | 250 | 140 |
| **x** | 220 | 130 | **ac** | 260 | 130 |

**Table 2: Measured torus sizes from the STXM XMCD projections in Figure S1.**

**Supplementary Note 2: Theoretical modelling:**

Theoretical modelling of the system was performed within the standard micromagnetic framework, assuming a Heisenberg exchange interaction and a bulk-type Dzyaloshinskii–Moriya interaction (DMI). These interactions are described by the Hamiltonian density:

$$H = A (\nabla \mathbf{m})^2 + D \mathbf{m} \cdot (\nabla \times \mathbf{m}),$$

where **m** is the unit vector along the magnetization, $A$ is the exchange stiffness, and $D$ is the bulk DMI constant. The micromagnetic formalism provides a continuum description and does not explicitly include finite-temperature effects; the states obtained therefore correspond to relaxed zero-temperature configurations of the Hamiltonian.

All simulations were performed in the absence of an external magnetic field, and thus Zeeman terms were neglected. The demagnetizing field was also neglected for the majority of the theoretical analysis; however, its role is examined in detail in a dedicated section of the Supplementary Notes below.

The material parameters used were: exchange stiffness $A = 5.7$ pJ/m, bulk DMI constant $D = 0.6$ mJ/m², and saturation magnetization $M_s = 230$ kA/m[1]. The Gilbert damping parameter was set to $\alpha = 1$ to facilitate numerical relaxation. The system is assumed to be free from defects, with homogenous material parameters, unless explicitly stated. Systems were initialized in a random state and allowed to relax over 20 ns. Figure S2 indicates that in the absence of a magnetic field, uniformly magnetized and randomized initial states all relax to the radial stripe state.

Finally, magnetocrystalline anisotropy terms were neglected. While such anisotropy can bias the orientation of helices in helimagnets it does not qualitatively modify the underlying ordering. Experimentally, we find that the radial stripe patterns follow the confined toroidal geometry, indicating that geometrical confinement dominates over crystalline anisotropy in our system. Anisotropy may contribute to secondary effects such as stripe straightening or defect pinning, but it is not expected to significantly affect the stability of the observed double-helix state. Nevertheless, anisotropy could in principle be exploited as a tuning parameter, for example to adjust the bulk helical period or to promote transitions between states with different winding numbers, as has been observed in bulk CoZnMn systems[2].

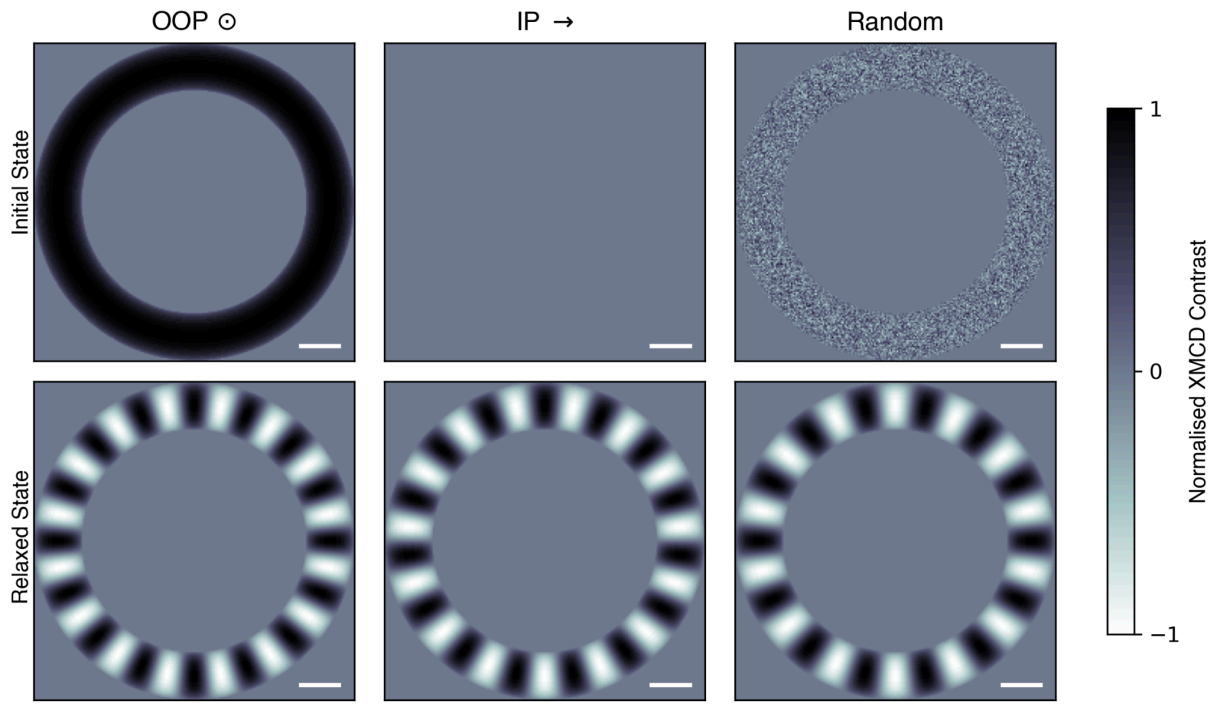

Figure S2: Relaxation of different initial magnetization states. Top row, XMCD projections of three simulated initial states: uniformly out-of-plane, uniformly in-plane, and a randomized configuration. Bottom row, corresponding relaxed states, each exhibiting the characteristic radial ordering. Scale bar indicates 100 nm.

**Supplementary Note 3: The role of magnetostatics:**
In the manuscript, we show that the direct exchange interaction and DMI are sufficient to recover the experimentally observed ordering in our system. Here, we explicitly address the role of the demagnetizing field in stabilizing the simulated states.

A simulated phase diagram tracking the role of the torus geometry, as in Figure 2, is shown in Figure S3; however, here the system is relaxed using a Hamiltonian containing the direct exchange interaction, DMI, and the magnetostatic interaction. We find that the phase diagram is essentially unchanged: the bulk multi-winding state still yields to the radial state around $r_{NW} = 1.2\lambda$. Only in the limiting regime of $r_{NW} < \lambda/2$ does the inclusion of magnetostatic qualitatively alter the phase diagram. In this regime, an in-plane vortex-ring state is stabilized, which minimizes the surface charges. A rendering and cross-sectional view of the magnetization for the vortex state are shown alongside the phase diagram in Figure S3. We therefore conclude that while magnetostatics can stabilize additional states at very small nanowire radii, they do not significantly affect the ordering in the experimentally relevant regime.

Notably, close to the transition between vortex and radial states we note a shift in the observed ordering of the radial state. Experimentally, this appears as the "zip-like" ordering seen both in Figure 4 and Figure S4. Figure S4 presents a direct comparison between representative experimental images of the standard double helix state, and the "zip-like" variant. (Fig. S4a,d), together with simulated projections of both projection types (Fig. S4b,e). Crucially, when we evaluate the associated field lines (Fig. S4c,f), we find that the two orderings are topologically equivalent. The zip-like pattern corresponds to an asymmetric double helix, in which the branches exhibit unequal sizes. This asymmetry originates from an increased in-plane canting of the bulk magnetisation, which lowers the demagnetising field energy in thinner samples, where surface charge contributions represent a larger fraction of the total energy. This constitutes an instance of spontaneous symmetry breaking, and can be viewed as an intermediate ordering between the standard double helix and vortex states.

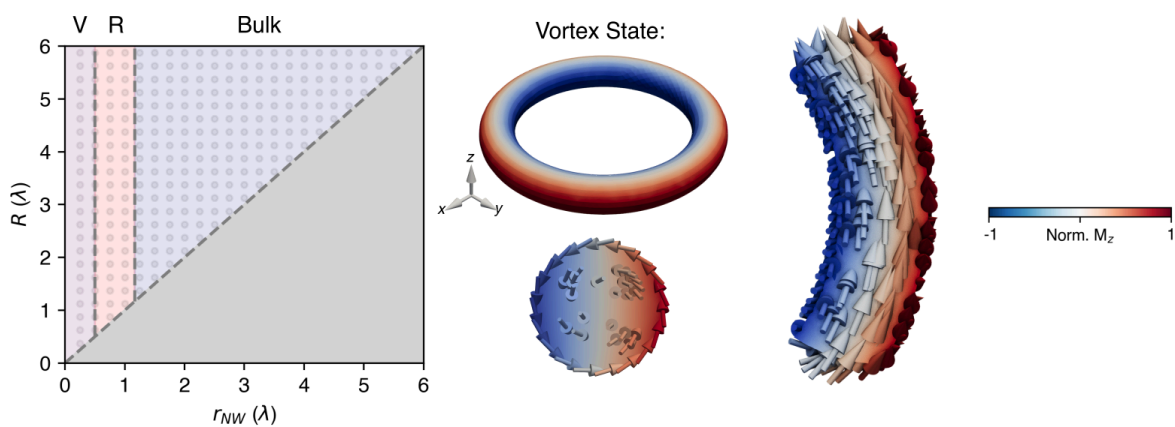

**Figure S3: Geometry dependence of the magnetic state within toroidal helimagnets including magnetostatics. Simulated geometric phase diagram for helimagnetic tori, where the Hamiltonian includes direct exchange interaction, Dzyaloshinskii–Moriya interaction (DMI), and magnetostatic interactions. 'V', 'R', and 'Bulk' represent the vortex, radial stripe, and bulk states, respectively. On the right, an isometric projection of a vortex state is shown, along with cross-sectional cuts through the x-y and x-z planes.**

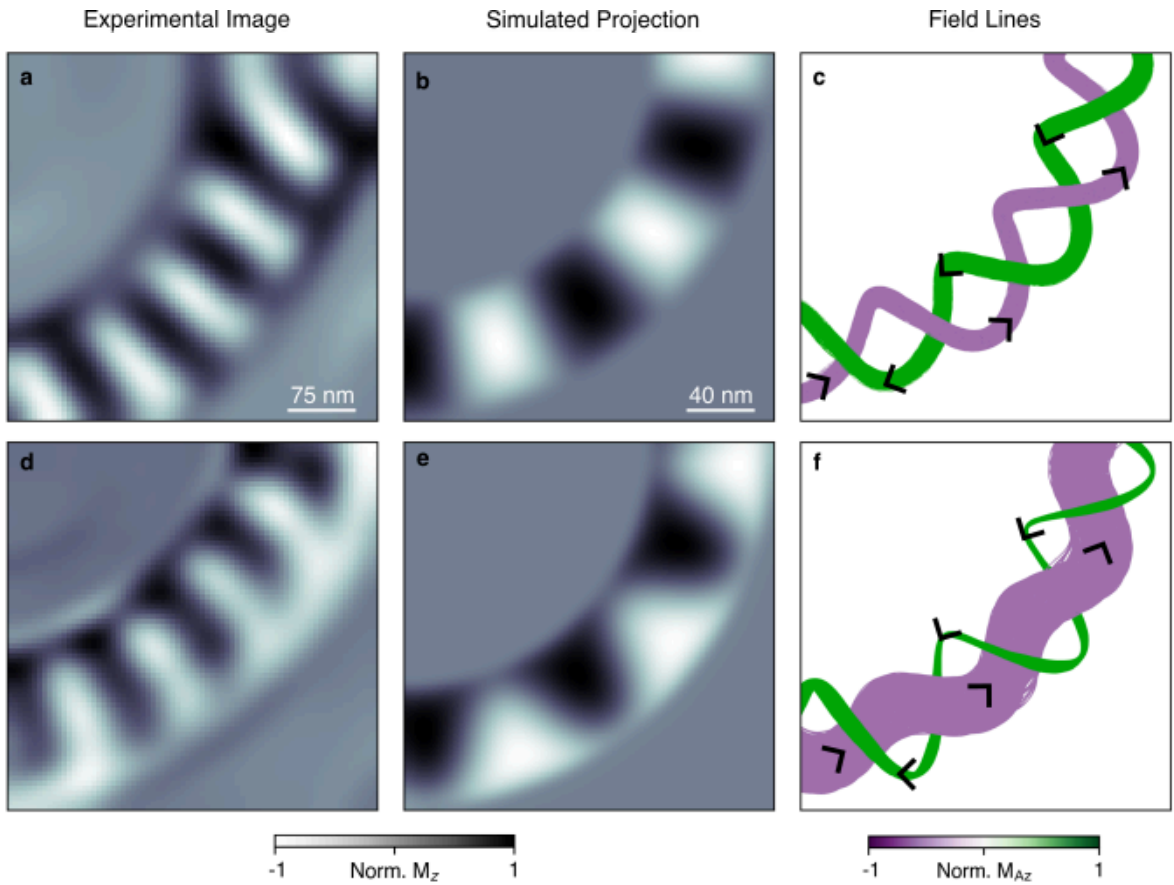

**Figure S4: Comparison of experimental and simulated projections of both the straight and zip-like ordering of the radially ordered states of the tori. (a) Experimental image of a torus ($R_T$ = 600 nm, $r_{NW}$ = 100 nm) displaying straight, separated stripes of white and black. (b) Corresponding simulation displaying separated stripes. (c) The surface field lines corresponding to the simulation in (b). (d–f) As in (a–c) but for the zip-like state.**

**Supplementary Note 4: The azimuthal component of the magnetization**

The azimuthal component of the magnetization was calculated by taking the cross product of the local magnetization vector, **m**, with the position vector, **r**, of **m**. The origin of the coordinate system was set to the center of the torus, and the **z**-components of **m** and **r** were excluded from the calculation to isolate the sense in which **m** circulates in the **x-y** plane.

**Supplementary Note 5: Sample Fabrication and Ga damage**

During focused ion beam (FIB) sample preparation, high energy gallium ions are focused onto the sample to locally mill material. The use of FIBs leads to the implantation of gallium ions into the sample, which can modify the crystallinity and structure of the underlying material, introducing disorder. Figure S1 shows the simulated damage profile of $Co_8Zn_9Mn_3$, as represented by the stopping distance of ions using the SRIM software package[3]. Ga+ ions accelerated at 30 kV and 16 kV, respectively, were simulated as incident to the surface normal, representing the maximal damage regime. We find that for 16 kV, the potential used to fabricate our samples, the damage layer is approximately 10 nm thick. This is consistent with previously reported values of damage layer thicknesses[4,5].

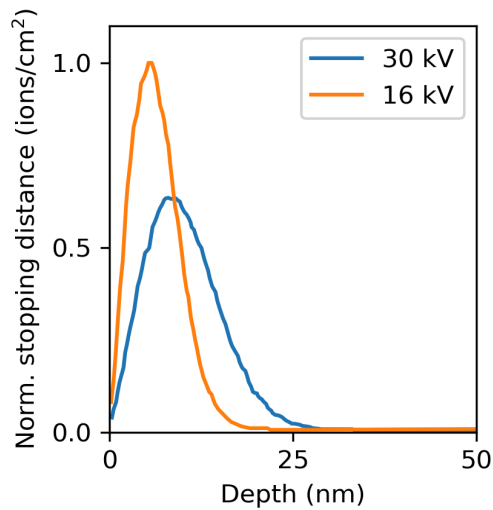

**Figure S5: Simulated damage profile as a function of Ga+ energy for Ga+ ions incident to the surface normal of $Co_8Zn_9Mn_3$. 16kV, and 30 kV potentials are shown in orange, and blue respectively.**

To investigate the impact of ion damage on the magnetization texture, we performed finite element micromagnetic simulations of a torus ($R_T$ = 360 nm, $r_{NW}$ = 120 nm), which includes a 10 nm thick shell of damaged material. Following the approach of Ref.[4], we approximated the properties of the damaged layer by using the bulk material parameters but with the Dzyaloshinskii-Moriya interaction (DMI) coupling constant $D$ suppressed. Figure S2 shows a series of finite element simulations with varying degrees of DMI suppression in the damaged layer. We observe that decreasing $D$ reduces the number of stripes that stabilize in the system. When the DMI is completely suppressed, as shown in the first column, the stripes no longer exhibit radial ordering; instead, their wave vectors tend to align along the radial direction of the torus ring. Examining the magnetization field lines, we find that the edge canting, associated with the double helix state, occurs in all damage regimes where radial ordering of stripes is present, even when the damage layer has completely suppressed DMI. We conclude that if the radial stripe pattern is present, one would expect to stabilize the double helical structure, even in the presence of damage.

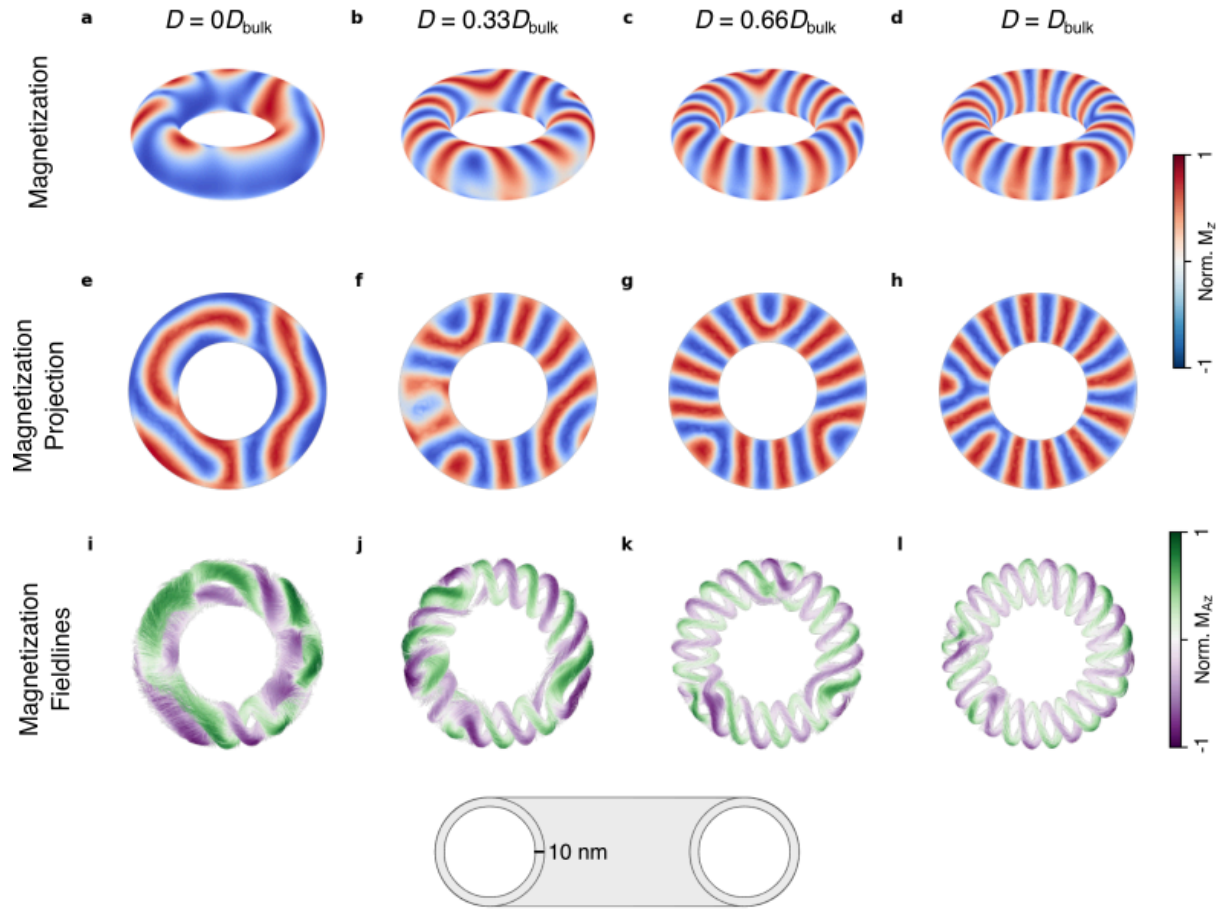

**Figure S6: Micromagnetic simulations of a torus with varying degrees of DMI suppression in the surface damage layer. (a–d) Simulated magnetic states as a function of DMI suppression within the surface damage layer, with the surface magnetisation indicated by the red-blue color scheme. The DMI coupling constant *D* is suppressed by 0%, 33%, 66%, and 100% of the bulk value, respectively. (e–h) A slice of the magnetization through the central plane of the torus corresponding to the states described above. A reduction in the number of stabilized stripes is apparent with decreasing *D*. In the limiting case where the DMI is completely suppressed, the radial stripe state no longer readily forms, except in a small region, highlighted by an arrow. (i–l) Magnetization field lines for the respective states. The surface canting associated with the double helix state is present in all damage regimes when the appropriate radial ordering of stripes is locally present.**